%% file: main.tex
\documentclass[sigplan,nonacm,balance=true]{acmart}
\usepackage{algorithm}
\usepackage{algorithmic}
\graphicspath{{figures/}}
\usepackage{textcomp}
\usepackage{tabularx}
\usepackage{colortbl}

\usepackage{tikz}
\usetikzlibrary{arrows.meta,calc,positioning}
\usepackage{listings}
\input{evaluation_macros}

\lstdefinestyle{motivation}{
    basicstyle=\ttfamily\footnotesize,
    breaklines=true,
    columns=fullflexible,
    keepspaces=true,
    showstringspaces=false,
    keywordstyle=\color{blue},
    commentstyle=\color{black}\bfseries,
    stringstyle=\color{red},
    frame=single,
    framesep=2pt,
    aboveskip=2pt,
    belowskip=2pt,
}

\begin{document}

\title{Automated Prefetching for Object Spatial Programming Using Temporal Trace Graphs}

\author{Baichuan Li}
\affiliation{
  \institution{University of Michigan}
  \city{Ann Arbor}
  \state{Michigan}
  \country{USA}
}
\email{patrli@umich.edu}

\author{Jayanaka Dantanarayana}
\affiliation{
  \institution{University of Michigan}
  \city{Ann Arbor}
  \state{Michigan}
  \country{USA}
}
\email{jayanaka@umich.edu}

\author{Savini Kashmira}
\affiliation{
  \institution{University of Michigan}
  \city{Ann Arbor}
  \state{Michigan}
  \country{USA}
}
\email{savinik@umich.edu}

\author{Lingjia Tang}
\affiliation{
  \institution{University of Michigan}
  \city{Ann Arbor}
  \state{Michigan}
  \country{USA}
}
\email{lingjia@umich.edu}

\author{Jason Mars}
\affiliation{
  \institution{University of Michigan}
  \city{Ann Arbor}
  \state{Michigan}
  \country{USA}
}
\email{profmars@umich.edu}

\renewcommand{\shortauthors}{Li et al.}

\begin{abstract}

% Dependent database accesses can place a sequence of application\allowbreak-database round trips on a request's critical path. Prefetching across these dependencies requires identifying future accesses before the application reaches them. We show how Object-Spatial Programming (OSP), which expresses computation as walkers over nodes connected by typed edges, exposes information for this purpose. Its traversal and ability-dispatch semantics specify how computation continues from newly discovered nodes, allowing local traversal rules to compose across walker steps. We introduce the Temporal Trace Graph (TTG), a bounded, request-specific prefetch plan. At walker spawn, the TTG Builder evaluates these rules over the live graph from the request's starting node to identify candidate objects. An asynchronous prefetcher loads these objects into the runtime cache while the walker executes. We implement the system in Jac and evaluate four applications with linear, multi-hop, recursive, and value-dependent traversals. The system reduces synchronous backing-store accesses and achieves up to $\headlineSpeedup\times$ end-to-end speedup over the stock runtime. We also characterize planning overhead and overfetch under value-dependent execution.

{\emergencystretch=1em
Dependent database accesses can serialize application--database round trips on a request's critical path. Planning across traversal steps requires knowing both which relationships a computation may follow and how discovered objects become subsequent execution contexts. We identify a programming-model contract that exposes this information through inspectable access paths and explicit \emph{traversal continuation}. We demonstrate this contract in Object-Spatial Programming (OSP), where walkers traverse a typed object graph and execute node-type-specific code. We introduce the Temporal Trace Graph (TTG), a bounded, request-specific graph of candidate objects and traversal-derived transitions. At walker spawn, the TTG Builder composes statically extracted access and continuation rules over the live graph from the request's starting node. An asynchronous prefetcher loads planned objects while the walker executes. We implement TTG in Jac/PostgreSQL and evaluate four applications covering linear, multi-hop, recursive, and value-dependent traversals. TTG improves end-to-end latency in three applications, achieving up to $\headlineSpeedup\times$ speedup over the stock runtime. Comparisons with manually staged prefetching characterize the benefits and costs of automatic cross-step planning. The evaluation also shows that planning overhead can exceed avoided demand latency and that value-dependent control can introduce bounded overfetch.\par}
\end{abstract}


\maketitle

\section{Introduction}

Database accesses often lie on the critical path of application requests. This cost is particularly high when consecutive accesses are dependent: the result of one request identifies the object needed by the next, forcing application–database round trips to occur sequentially. Although individual queries may be fast, repeatedly crossing the application–storage boundary can dominate request latency~\cite{lookingglass2}. Prefetching can remove these stalls, but only if the runtime can determine which objects will be needed before application execution reaches their accesses.

DBridge~\cite{dbridge-prefetch} and CAPre~\cite{capre} identify future database accesses through program analysis. AutoFetch~\cite{autofetch} derives prefetch decisions from traversal profiles, while SeLeP~\cite{selep} uses access-history prediction. Sloth~\cite{sloth} batches queries during application execution. Table~\ref{tab:information-sources} summarizes representative mechanisms. We focus on the information needed to plan across dependent traversal steps. An access path describes how a computation reaches related objects. Composing such paths across steps also requires knowing how computation may continue from those objects: which become subsequent execution contexts, and which accesses may occur there. This connection between object discovery and subsequent computation allows local access rules to compose into a candidate plan rooted at the current request.

We call this connection \emph{traversal continuation}. In conventional object-oriented programming (OOP), recovering this connection statically can be difficult because it is encoded in general-purpose control and data flow. For example, a breadth-first traversal places newly discovered objects in a queue and processes them in later loop iterations. An analyzer must track these objects through the queue and identify the accesses performed from them when they are processed. This reasoning may involve aliases, container operations, and method calls across application and library code. An analyzer must reconstruct these connections to compose local association accesses into a description of the overall traversal.

We identify a programming-model contract for request-specific read planning: inspectable access paths paired with explicit traversal continuation. Access paths expose the persistent relationships a computation may follow; continuation rules identify which access targets may become subsequent execution contexts and which computations may execute there. When these rules can be resolved over live persistent relationships from the request's starting object, a runtime can compose them into a bounded candidate plan before traversal execution.

Object-Spatial Programming (OSP)~\cite{mars2025osp} provides one instantiation of this contract. It extends OOP with typed nodes and edges for data and stateful walkers for computation. Typed traversal expressions expose access paths. A \texttt{visit} statement schedules target nodes as future walker execution contexts, while ability-dispatch rules specify which computations may execute when those nodes are reached. Together, these constructs make local access rules composable across walker steps.

\begin{table}[!ht]
\centering
\caption{Access information and fetching mechanisms in representative systems.}
\label{tab:information-sources}
\Description{Four systems are compared by the source of their access structure and their fetching mechanism. CAPre infers paths from OOP code and generates prefetch code that traverses runtime objects. AutoFetch uses observed traversals to augment queries with fetch specifications. SeLeP learns sequences of data encodings to predict partitions for subsequent queries. The highlighted TTG row uses explicit access and continuation rules to build a bounded candidate plan over the live graph before traversal execution. Value-dependent control can cause TTG to overfetch.}
\small
\setlength{\tabcolsep}{4pt}
\renewcommand{\arraystretch}{1.15}
\begin{tabularx}{\columnwidth}{>{\raggedright\arraybackslash}p{0.22\columnwidth}>{\raggedright\arraybackslash}X>{\raggedright\arraybackslash}X}
\toprule
\textbf{System} & \textbf{Access structure} & \textbf{Fetch mechanism} \\
\midrule
CAPre~\cite{capre} & Paths inferred from OOP code & Generated prefetch code traverses runtime objects \\
\addlinespace[3pt]
AutoFetch~\cite{autofetch} & Previously observed traversals & Fetch specifications augment queries \\
\addlinespace[3pt]
SeLeP~\cite{selep} & Learned sequences of data encodings & Predicts partitions for subsequent queries \\
\midrule
\rowcolor{blue!6}
\textbf{TTG (this work)} & \textbf{Explicit access and continuation rules} & Bounded plan over the live graph \textbf{before traversal} \\
\bottomrule
\end{tabularx}
\par\smallskip
\begin{minipage}{\columnwidth}
\footnotesize
TTG requires inspectable access paths and explicit continuation rules; value-dependent control may cause overfetch.
\end{minipage}
\end{table}

We exploit this contract with the \textbf{Temporal Trace Graph (TTG)}, a bounded, request-specific representation of candidate persistent accesses. The TTG Builder composes access and continuation rules over the live graph from the request's starting object, identifying candidate objects before traversal execution. In our Jac implementation, a static analyzer extracts these rules from walker abilities and their helper functions, and the Builder constructs the plan at walker spawn. An asynchronous prefetcher loads planned objects while the traversal executes, allowing later demand accesses to reuse both prefetched object data and previously resolved topology.

We implement TTG in the Jac OSP runtime \cite{jaseci} with PostgreSQL and evaluate four applications covering linear, multi-hop, recursive, and value-dependent traversals. TTG reduces synchronous backing-store accesses and improves end-to-end latency by up to $\headlineSpeedup\times$. The evaluation also exposes the limits of the TTG: value-dependent pruning in user programs can conservatively cause TTG to introduce unused prefetches, and TTG is not profitable when its overhead exceeds the demand stalls it removes. These results show both the opportunity created by explicit traversal semantics and the conditions under which the opportunity translates into performance.

This paper makes three contributions. First, we identify a programming-model contract for request-specific read planning: inspectable access paths paired with explicit traversal continuation, with OSP providing one instantiation. Second, we design the TTG Builder to compose statically extracted access and continuation rules over the live graph at request spawn, producing bounded candidate plans across future traversal steps. Third, we implement TTG in Jac/PostgreSQL and evaluate four traversal workloads, demonstrating up to $\headlineSpeedup\times$ end-to-end speedup over the stock runtime while characterizing planning overhead, cache behavior under value-dependent control, and trade-offs relative to manually staged prefetching.

\section{Background}
\label{sec:bg}

\subsection{Object-Spatial Programming}
\label{sec:bg:jac}

Object-Spatial Programming adds three class kinds to object-oriented programming: \texttt{node}, \texttt{edge}, and \texttt{walker}. A \texttt{node} represents a data-carrying vertex; an \texttt{edge} represents a typed relationship between nodes; and a \texttt{walker} carries application logic and state, executing that logic as it visits nodes. Each declaration defines a class, with \texttt{has} fields specifying instance state. Node and edge instances together form the \emph{OSP graph}, over which walkers execute.

Listing~\ref{lst:osp-types} declares two node types and one edge type.

\begin{lstlisting}[language=Python, caption={Node and edge declarations.}, label={lst:osp-types}]
node User { has name: str; }
node Tweet { has text: str; }
edge Posted;
\end{lstlisting}

A walker expresses its application logic through methods called abilities. An ability can be triggered when the walker enters or exits a node of a specified type. Inside an ability, \texttt{here} refers to the current node and \texttt{self} refers to the walker instance. Abilities can read or update their state and issue \texttt{visit} statements to schedule nodes for subsequent visits. When the walker enters a scheduled node, that node becomes \texttt{here}, and the runtime invokes the matching entry abilities.

Traversal expressions select related nodes through typed edges, with arrow notation specifying the direction. For example, \texttt{[->:\allowbreak Follows:\allowbreak ->->:\allowbreak Posted:\allowbreak ->]} follows outgoing \texttt{Follows} edges from the current user and then outgoing \texttt{Posted} edges from those users, yielding their tweets. A \texttt{visit} statement schedules the resulting nodes for subsequent walker visits. In this example, the tweets are scheduled; the intermediate users participate in path resolution but are not themselves scheduled by this statement. Traversal expressions can also be used to access related objects within an ability without scheduling further visits.

Listing~\ref{lst:osp-walker} defines a walker that gathers tweets posted by the users a given user follows.

\begin{lstlisting}[language=Python, caption={A minimal OSP walker.}, label={lst:osp-walker}]
walker Timeline {
    has feed: list[Tweet] = [];

    can gather with User entry {
        visit [->:Follows:->->:Posted:->];
    }
    can collect with Tweet entry {
        self.feed.append(here);
    }
}
\end{lstlisting}

A running application launches a walker with \texttt{spawn}:

\begin{lstlisting}[language=Python]
result = root spawn Timeline();
\end{lstlisting}

When \texttt{Timeline} starts at \texttt{u}, the runtime invokes \texttt{gather}, whose \texttt{visit} statement schedules the tweets selected by the two-hop traversal. At each scheduled \texttt{Tweet}, the runtime invokes \texttt{collect}, with \texttt{here} referring to that tweet. The ability appends the tweet to the walker's \texttt{feed} field. Because \texttt{collect} schedules no further visits, traversal finishes after the queued tweets have been processed.

Figure~\ref{fig:jdrive-code-motivation} contrasts the continuation mechanisms in OOP and OSP using a folder-tree traversal. In the OOP version, recognizing repeated descent requires connecting \texttt{folder.\allowbreak children} to the queue update, the subsequent dequeue, and execution of the loop body on the dequeued folder. The schema association alone does not express this continuation. In OSP, the \texttt{Contains} visit selects child folders, and the \texttt{Folder} entry trigger specifies that the same ability may execute at each target and issue further visits. OOP programs can expose equivalent structure through a traversal API or recover it through program analysis.

\begin{figure}[t]
\centering
\begin{minipage}[t]{0.48\columnwidth}
\centering\footnotesize\textbf{Python ORM}
\begin{lstlisting}[style=motivation,language=Python]
from collections import deque
def visible_tree(root):
    frontier, visible = deque([root]), []
    while frontier:
        folder = frontier.popleft()
        if folder.trashed:
            continue
        visible.append(folder)
        # Continuation encoded through worklist.
        frontier.extend(folder.children)
    return visible
\end{lstlisting}
\end{minipage}\hfill
\begin{minipage}[t]{0.48\columnwidth}
\centering\footnotesize\textbf{OSP walker}
\begin{lstlisting}[style=motivation,language=Python]
walker VisibleFolderTree {
    has visible: list[Folder] = [];
    can collect with Folder entry {
        if here.trashed { # VALUE: prune
            return;
        }
        self.visible.append(here);
        # Explicit traversal continuation.
        visit [->:Contains:->[?:Folder]];
    }
}
\end{lstlisting}
\end{minipage}
\caption{Simplified folder-tree traversals in Python ORM (OOP) and OSP. Here, \texttt{folder.children} returns child folders. The ORM version expresses repeated descent through queue operations and loop control; OSP expresses it through \texttt{visit} and the \texttt{Folder} entry trigger.}
\label{fig:jdrive-code-motivation}
\end{figure}

Without TTG, data access in the Jac runtime is demand-driven. Traversal involves two kinds of operations: \emph{topology resolution}, which identifies nodes reached by a traversal expression, and \emph{object materialization}, which makes their contents available to application code. These operations reuse cached topology and object data when available. When required data must be fetched from the backing store, the runtime issues a synchronous read and blocks the walker until the result arrives. The runtime can batch reads within a traversal operation, but initiates them only when execution reaches that operation.

\subsection{Problem statement}
\label{sec:bg:problem}

We now state the problem TTG solves. For a given request, let $W$ be its OSP walker, $n_0$ its spawn node, and $G$ the live node-edge graph at spawn time. Let $A(W,n_0,G) \subseteq V(G)$ denote the set of nodes actually read by the request, and let $L$ be a user-configured bound on the number of nodes in the prefetch plan. The planning problem is to construct a candidate set $P \subseteq V(G)$ before $W$ begins executing, using information available at spawn time. This set guides subsequent prefetching into the runtime cache, with the following coverage objective and size constraint.

\textbf{Coverage objective.} Within the budget $L$, the goal is to select $P$ to cover as many nodes in $A(W,n_0,G)$ as possible, using only information available at spawn time.

\textbf{Boundedness.} $|P| \leq L$.

{\emergencystretch=1em
The set $P$ contains candidate nodes for prefetching. Value-dependent pruning during walker execution may leave planned nodes unused, while the size bound may leave some requested nodes outside $P$. During execution, data unavailable in the runtime cache is fetched through the ordinary demand-read path. Section~\ref{sec:impl:builder} describes how the TTG Builder constructs $P$; Section~\ref{sec:impl:prefetcher} explains how the Prefetcher loads the selected nodes into the runtime cache.\par}

\textbf{Temporal Trace Graph.} A TTG is a bounded, request-specific directed graph $T=(P,E,\ell)$, rooted at the spawn node $n_0$. Its vertices $P$ are candidate objects for prefetching. An edge $(u,v)\in E$ records that applying a traversal rule at $u$ identifies $v$ as a candidate; the rule may resolve a multi-hop path in the underlying object graph. Targets scheduled by \texttt{visit} are eligible for further expansion. Each vertex carries a level $\ell(v)$ corresponding to its first discovery during breadth-first expansion from $n_0$. These levels capture logical traversal progression rather than execution timestamps.

\section{Implementation}

\subsection{Architecture Overview}
\label{sec:impl:arch}

\begin{figure*}[t]
\centering
\includegraphics[width=\linewidth]{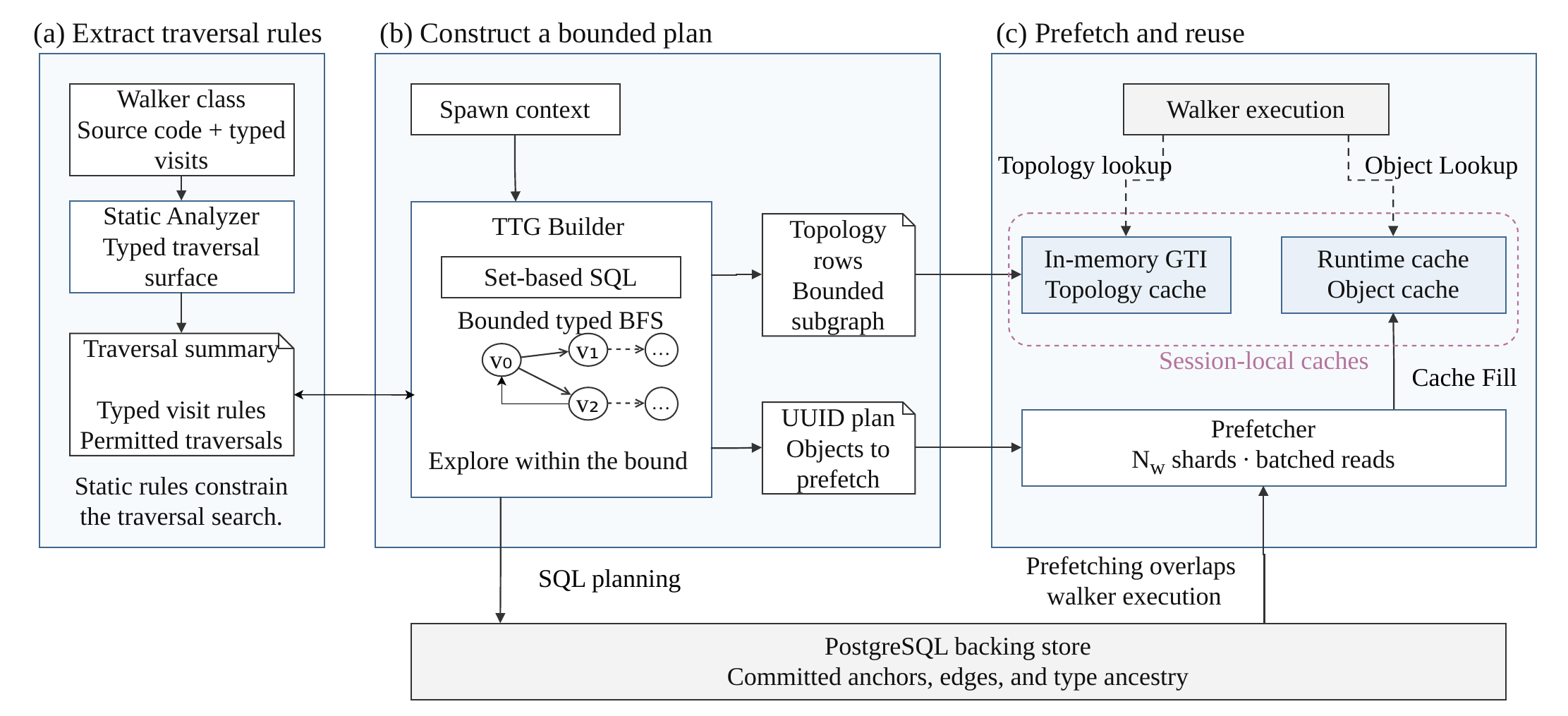}
\caption{Architecture of the TTG system. The Static Analyzer extracts traversal rules from the walker class. Using these rules and the spawn context, the TTG Builder queries PostgreSQL to construct a bounded prefetch plan and populate the session-local topology cache (GTI). The Prefetcher loads planned object data into the runtime cache while the walker executes. Demand accesses reuse cached object data and topology.}
\label{fig:architecture}
\end{figure*}

\begin{figure}[t]
    \centering
    \includegraphics[width=\linewidth]{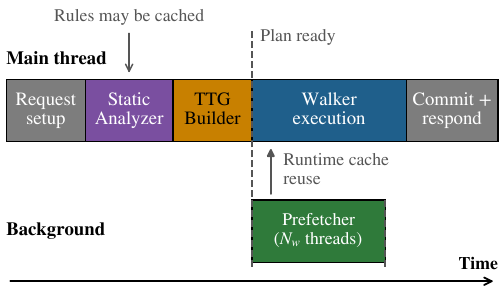}
    \caption{Runtime timeline of a Jac OSP walker with TTG prefetcher enabled.}
    \label{fig:timeline}
\end{figure}

% Figure~\ref{fig:architecture} separates the system into three layers: the walker runtime, TTG, and PostgreSQL, with the walker class shown as a program input outside the runtime layer. The walker class feeds TTG's Static Analyzer, while the runtime supplies the spawn root and plan bound $L$ to the TTG Builder. TTG then runs a bounded planning query over PostgreSQL's committed graph rows and emits two flows: a BFS-ordered prefetch set $P$ with $|P| \leq L$ for the Prefetcher, and topology rows for the session-local in-memory GTI. The Prefetcher (\S\ref{sec:impl:prefetcher}) batch-reads $P$ into the shared runtime cache while ordinary walker execution continues through the same cache-backed demand-read path. Typed graph resolves can consume the GTI when TTG has warmed the relevant hops; residual cache misses continue to PostgreSQL through the existing read path. The rest of this section describes each component in turn.

Figure~\ref{fig:architecture} shows how the TTG system integrates with the Jac runtime and PostgreSQL. The Static Analyzer extracts typed traversal rules from the walker class. At walker spawn, the TTG Builder evaluates these rules over the live graph from the spawn node through a synchronous planning query in PostgreSQL. It constructs a TTG containing at most $L$ candidate node identifiers in breadth-first order. The same query returns topology rows that populate the \emph{Graph Topology Index (GTI)}, a session-local topology cache. Once planning completes, the Prefetcher batch-loads the planned object data into the runtime cache while the walker executes. The walker's ordinary demand-access path reuses cached object data and resolves traversal expressions using the cached topology when available.

The planning query returns node identifiers and topology rows, separating topology resolution from object loading. This allows the runtime to populate the topology cache and begin walker execution as soon as planning completes, while the Prefetcher loads object data through parallel batched reads. The walker's demand-access path reuses the cached topology and any object data already available in the runtime cache.

Figure~\ref{fig:timeline} illustrates the runtime of a general walker when the TTG prefetching is enabled.

\subsection{Static Analyzer}
\label{sec:impl:analyzer}

\textbf{Problem.} To construct a TTG at spawn time, the Builder needs to know which related objects each ability may access and how traversal can continue from the nodes it schedules for visits. The Static Analyzer must extract the supported typed access paths together with their ability triggers and distinguish ordinary association reads from paths used in \texttt{visit} statements. This information enables the Builder to compose access rules across walker steps.

\textbf{Idea.} OSP exposes access paths through traversal expressions and traversal continuation through \texttt{visit} statements and ability triggers. The analyzer walks each ability's AST and follows statically resolvable calls into helper functions. It collects supported traversal expressions, associates them with the enclosing ability's trigger types, and records whether they are used to schedule further visits. When an expression's edge type cannot be resolved statically, the analyzer records the generic edge class from which OSP edges inherit, allowing the Builder to conservatively expand across its subtypes during planning.

\textbf{Example.} For the \texttt{gather} ability in Listing~\ref{lst:osp-walker}, the analyzer records the \texttt{User} entry trigger and the two-hop path (\texttt{Follows}, \texttt{Posted}), whose \texttt{Tweet} endpoints are scheduled for visits. The \texttt{collect} ability has a \texttt{Tweet} entry trigger and no further traversal expressions. The path's intermediate users are marked \emph{skip-eligible}: they participate in topology resolution, but their fields are not read by the application's logic in this example. The Builder can therefore resolve the full path while selecting only its endpoints for prefetching (\S\ref{sec:impl:builder}).

\textbf{Implementation.} The Jac runtime caches the Static Analyzer's extracted traversal rules on the loaded walker class. Subsequent requests using that class in the same runtime process reuse the cached rules, avoiding repeated AST analysis. The Builder still constructs a request-specific TTG at each spawn by resolving these rules over the live graph from the request's spawn node.

% \begin{lstlisting}[language=Python, caption={A minimal OSP walker with a multi-hop traversal.}, label={lst:timeline-walker}]
% walker Timeline {
%   has feed: list[Tweet] = [];

%   can gather with User entry {
%     visit [->:Follows:->->:Posted:->];
%   }

%   can collect with Tweet entry {
%     self.feed.append(here);
%   }
% }
% \end{lstlisting}

\subsection{TTG Builder}
\label{sec:impl:builder}

\textbf{Problem.} The Static Analyzer provides typed access paths associated with ability triggers. To construct a request-specific TTG, the Builder must resolve these paths over the live graph from the spawn node to obtain concrete candidate node identifiers. For nodes reached through visit paths, it must determine which ability rules may apply and use those rules to derive subsequent accesses.

\textbf{Challenge.} The number of objects reachable through the traversal rules depends on the stored graph. Deep paths and broad frontiers can produce large candidate sets, while cycles can cause repeated discovery of the same nodes. The Builder must construct a candidate plan of at most $L$ nodes from this data-dependent reachable set.

\textbf{Idea.} The Builder expands the walker's traversal rules breadth-first from the spawn node. At each frontier node, it applies the rules associated with matching ability triggers to identify candidate objects. Targets scheduled by \texttt{visit} paths form the frontier for subsequent walker-step expansion. The plan retains candidates in breadth-first order up to the configured limit $L$. In the PostgreSQL-backed implementation, a recursive SQL query performs this expansion and returns both the plan UUIDs and topology rows used to populate the session-local GTI.

\textbf{Algorithm.} Algorithm~\ref{alg:builder} maintains a set $P$ of distinct candidate UUIDs and a frontier of nodes to expand. For each frontier node, \textsc{MatchingRules} selects the applicable ability rules, and \textsc{ResolvePath} resolves each complete typed path to its target UUIDs. These targets become prefetch candidates; targets of \texttt{visit} paths are also eligible for subsequent expansion. New candidates are deduplicated against $P$ and truncated in topology order to fit the remaining budget. The next frontier contains admitted visit targets that have not yet been expanded, including nodes previously admitted through ordinary reads. Construction stops when the frontier is empty or $|P| = L$.

% \begin{algorithm}[t]
% \caption{TTG Builder: bounded typed BFS from the spawn node.}
% \label{alg:builder}
% \begin{algorithmic}[1]
% \REQUIRE walker class $W$, spawn node $n_0$, limit $L$, Analyzer records
% \ENSURE level-indexed list of node UUIDs, $|plan| \leq L$
% \STATE $frontier \gets \{n_0\}$
% \STATE $plan \gets [\{n_0\}]$;\ $total \gets 1$
% \WHILE{$frontier \neq \emptyset$ \textbf{and} $total < L$}
%     \STATE $next \gets \emptyset$
%     \FOR{each node $u \in frontier$}
%         \STATE $rec \gets$ Analyzer record for class$(u)$ under $W$
%         \IF{$rec = \emptyset$}
%             \STATE \textbf{continue}
%         \ENDIF
%         \FOR{each edge type $e \in rec.edges$}
%             \STATE $next \gets next \cup$ \textsc{ExpandOne}($u, e, rec$)
%         \ENDFOR
%     \ENDFOR
%     \IF{$total + |next| > L$}
%         \STATE truncate $next$ to $L - total$ items in topology order
%     \ENDIF
%     \STATE $plan.\textsc{Append}(next)$;\ $total \gets total + |next|$
%     \STATE $frontier \gets next$
% \ENDWHILE
% \STATE \textbf{return} $plan$
% \end{algorithmic}
% \end{algorithm}

\begin{algorithm}[t]
\caption{TTG Builder: logical breadth-first expansion with a node budget.}
\label{alg:builder}
\begin{algorithmic}[1]
\REQUIRE graph $G$, spawn node $n_0$, limit $L \geq 1$, Analyzer rules $R$ for walker $W$
\ENSURE level-indexed plan containing at most $L$ distinct node UUIDs
\STATE $P \gets \{n_0\}$; $plan \gets [\{n_0\}]$
\STATE $frontier \gets \{n_0\}$; $expanded \gets \emptyset$
\WHILE{$frontier \neq \emptyset$ \textbf{and} $|P| < L$}
\STATE $candidates \gets \emptyset$; $visits \gets \emptyset$
\STATE $expanded \gets expanded \cup frontier$
\FOR{each node $u \in frontier$}
\FOR{each rule $r \in \textsc{MatchingRules}(R,u)$}
\STATE $targets \gets \textsc{ResolvePath}(G,u,r)$
\STATE $candidates \gets candidates \cup targets$
\IF{$r.is_visit$}
\STATE $visits \gets visits \cup targets$
\ENDIF
\ENDFOR
\ENDFOR
\STATE $candidates \gets candidates \setminus P$
\IF{$|candidates| > L - |P|$}
\STATE truncate $candidates$ to $L - |P|$ nodes in topology order
\ENDIF
\STATE $plan.\textsc{Append}(candidates)$
\STATE $P \gets P \cup candidates$
\STATE $frontier \gets (visits \cap P) \setminus expanded$
\ENDWHILE
\STATE \textbf{return} $plan$
\end{algorithmic}
\end{algorithm}

\textbf{Implementation.} For PostgreSQL, the Builder constructs the planning query in two stages. It first compiles each transition rule into a SQL fragment that resolves the rule's typed access path. It then composes these fragments into a recursive common table expression (CTE), seeded by the request's spawn node, to expand accesses across walker steps. The resulting single query executes in PostgreSQL and returns at most $L$ candidate node UUIDs together with topology rows. The Builder directly lowers the TTG into a deduplicated, BFS-ordered UUID list for the Prefetcher (\S\ref{sec:impl:prefetcher}), without materializing a separate TTG adjacency structure.

The Builder uses the returned topology rows to populate the GTI, a session-local cache of resolved graph relationships. The runtime consults these entries when evaluating typed traversal expressions; required topology that is not cached is resolved through the existing backing-store path. Object data is loaded separately by the Prefetcher, allowing walker execution to begin with the resolved topology available while object prefetching proceeds asynchronously.

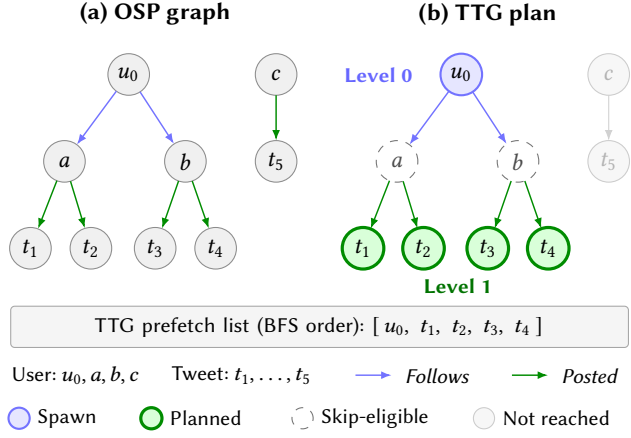
\begin{figure}[t]
\centering
\input{figures/plan_example.tikz}
\caption{Algorithm~\ref{alg:builder} applied to the live OSP graph for the Timeline walker of Listing~\ref{lst:osp-walker}, showing typed edges and chain compression. (a): an excerpt of the graph including spawn User $u_0$, two Users $u_0$ Follows, their Tweets, and one User $u_0$ does not Follow (User $c$) with $c$'s Tweet $t_5$. (b): the plan (Level 0 = $u_0$; Level 1 = the four Tweets reached via the \texttt{Follows}/\texttt{Posted} chain), with Follows-target Users drawn dashed because the Analyzer flagged them skip-eligible (traversed for chain resolution but not prefetched), and $c$ / $t_5$ faded because $u_0$'s typed visits never reach them. The boxed annotation is the actual data structure the Prefetcher consumes: an ordered UUID list in BFS order.}
\label{fig:plan-example}
\end{figure}

Figure~\ref{fig:plan-example} applies the traversal rules to the Timeline walker in Listing~\ref{lst:osp-walker}. The two-hop \texttt{Follows}/\texttt{Posted} chain contributes its tweet endpoints to the plan in a single level transition. Intermediate users $a$ and $b$ participate in path resolution but are omitted from the prefetch candidates because the example does not read their fields. The boxed annotation shows the resulting candidate UUIDs in BFS order.

\subsection{Prefetcher}
\label{sec:impl:prefetcher}

\textbf{Problem.} A TTG specifies candidate node UUIDs in BFS order. The Prefetcher's task is to load the corresponding object data into the runtime cache while the walker executes, with the goal of making that data available before the walker demands it.

\textbf{Idea.} The Static Analyzer and TTG Builder run sequentially on the request's main thread. Once planning completes, the Prefetcher uses $N_w$ background workers to fetch the planned object data while the main thread executes the walker. The walker's ordinary demand-read path reuses prefetched data already available in the shared runtime cache.
Figure~\ref{fig:timeline} illustrates the synchronous planning phase and the subsequent overlap between walker execution and asynchronous prefetching.

\textbf{Algorithm.} The Prefetcher assigns the plan's UUIDs in BFS order to $N_w$ workers using round-robin partitioning. Each worker issues one batched storage read for its assigned UUIDs and stores the returned raw documents in the runtime cache, keyed by UUID. The demand-read path can reuse these documents once they have been cached.

\textbf{Implementation.} The evaluated runtime uses CPython 3.14's free-threaded build (PEP~703). Prefetch workers fetch documents by UUID and store them as raw Python dictionaries in the shared runtime cache, leaving the runtime's write-set unchanged. The ordinary demand-read path materializes Jac objects from these documents as needed and reuses objects that are already materialized.

\section{Evaluation}
\label{sec:eval}

We evaluate the TTG system in terms of request performance and cache behavior. End-to-end latency measures its effect on request execution, while backing-store request counts characterize storage-access activity. Runtime-cache hit rate captures data availability at demand time; cache-derived diagnostics relate distinct-object cache hits to reported plan size. We sweep the prefetch limit to examine how plan size affects latency, planning overhead, and cache reuse.

\subsection{Experimental Setup}
\label{sec:eval:method}

\textbf{Workload.} We evaluate four OSP applications that exercise different traversal patterns.

\emph{LinkedList} contains a chain of \texttt{Item} nodes connected by \texttt{Next} edges. Its \texttt{Traverse} walker reports the current item's value and schedules a visit to the next item, exercising repeated single-hop traversal along the chain.

{\emergencystretch=1em
\emph{LittleX} is a social network modeled on Twitter. Its \texttt{load\_feed} walker gathers and sorts tweets posted by the current user and the users they follow. It accesses the current user's tweets through \texttt{Posted} edges and followed users' tweets through the \texttt{Follows}/\texttt{Posted} path illustrated in Listing~\ref{lst:osp-walker}. The database used in this test has 25,000 users and 5,000,000 Tweets in total.\par}

\emph{Jacord} is a Discord/Slack-like instant messenger. Its walker, called \texttt{load\_channel}, gathers channel messages and recursively follows incoming \texttt{ReplyTo} edges to collect replies. For each message, it also reads the author's username and counts direct replies. The database used in this test has 1,120 channels with 3,598,280 messages in total.

\emph{JDrive} is a cloud-drive application organized as a folder tree. Its \texttt{Visible\allowbreak Folder\allowbreak Tree} walker follows \texttt{Contains} edges to child folders whose \texttt{trashed} flag is false, pruning the subtrees rooted at trashed children. This workload exercises value-dependent recursive traversal. The database used in this test has 361 root folders and 12,328,511 folders in total.

\begin{figure*}[t]
\centering
\input{figures/workload_graphs.tikz}
\caption{OSP graph shapes for the four evaluation workloads. Each panel shows a small concrete instance so the node classes, edge types, and characteristic branching are visible at a glance. The walker's spawn node is drawn in blue in every panel; ellipses mark continuations, and the dashed JDrive branch shows a subtree pruned by the runtime \texttt{trashed} predicate. Actual working-set sizes in the evaluation are far larger than the instances drawn here.}
\label{fig:workload-graphs}
\end{figure*}
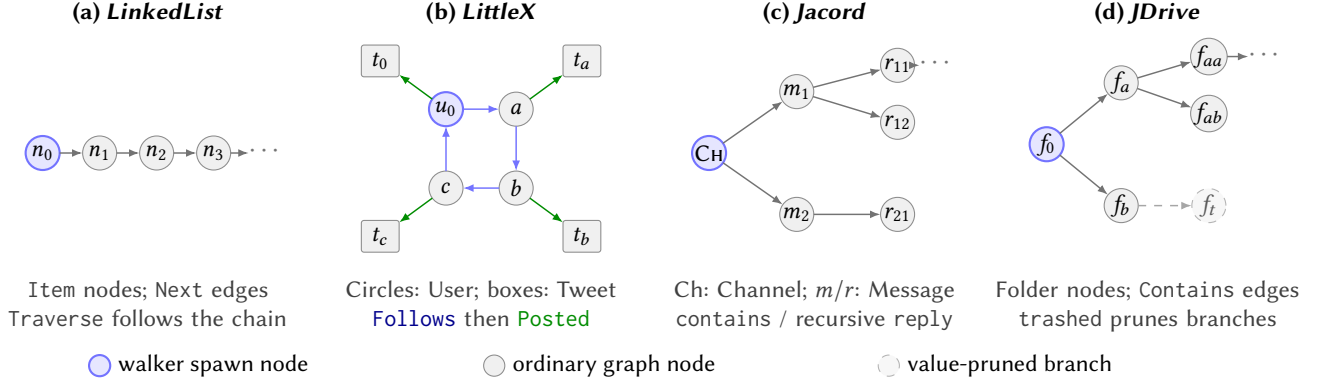

Figure~\ref{fig:workload-graphs} illustrates the graph structures and traversal patterns of the four workloads using small example instances.

\textbf{Hardware and deployment.} The prefetch-limit sweeps use two machines on the same university LAN. Both machines have an Intel Core i9-10850K CPU (10 cores, 20 hardware threads) and 62~GiB of RAM. Machine 1 runs Pop!\_OS 22.04 with Linux 6.2.6; Machine 2 runs Ubuntu 20.04 with Linux 5.15 and Postgres 17.11. Machine 1 hosts the measurement harness and Jac server, while Machine 2 hosts PostgreSQL. The Jac server and PostgreSQL run in Docker containers and communicate over TCP. Static analysis and planning-query construction run on Machine 1 alongside the Prefetcher and walker; PostgreSQL on Machine 2 executes the planning query and storage-read queries.

\textbf{Baselines and policies.} Our primary baseline is the stock Jac runtime with TTG disabled. It retains the runtime's standard caching and batching mechanisms and resolves topology and materializes objects through the ordinary demand-access path.

{\emergencystretch=1em
We use \emph{Manual staged prefetch} (MSP) to characterize manually optimized staged prefetching without an explicit traversal-continuation interface. MSP manually implements CAPre's association-path prefetching~\cite{capre} together with DBridge's early asynchronous query submission, dependency chaining, and compatible batching~\cite{dbridge,dbridge-prefetch}. For each walker ability, we manually encode the association paths accessed by its body and called helper functions. Background helpers resolve these paths from the current node and available parameters, then load the resulting objects through Jac's standard object-loading functions and runtime cache. Helpers are submitted at points corresponding to ability entry; JDrive additionally submits child helpers when children are enqueued. Object loads are chained to the topology queries that produce their identifiers, while compatible independent accesses share topology queries and object-load batches. Traversal continuation remains encoded in application control flow and helper-submission logic, rather than exposed as rules that a planner can compose across steps before traversal execution.\par}

The backing-store request-count comparison in Figure~\ref{fig:db-requests-all} additionally includes \emph{Oracle}, which is supplied with the request's actual node read set and prefetches exactly those nodes. This provides a reference with exact knowledge of the requested objects.

We include SeLeP~\cite{selep} as a history-based block-prefetching baseline (\S\ref{sec:eval:selep}). SeLeP runs as a sidecar alongside the Jac runtime. During request execution, Jac emits SQL trace events that the sidecar consumes incrementally to predict subsequent block accesses. The sidecar issues \texttt{pg\_prewarm} calls to PostgreSQL to prefetch the selected blocks, while Jac continues executing through its ordinary demand-access path.

\textbf{Metrics.} For each measured request, we record \texttt{e2e\_ms}, runtime component timings, cache statistics, access logs, and execution profiles. For TTG, the reported latency includes synchronous plan construction and walker execution, including waits for data. We analyze access logs after execution to compute runtime-cache hit rate and cache-derived diagnostics. Backing-store request counts are derived from profiled PostgreSQL backend calls. Instrumentation performed during request execution is included in the reported latency.

% \subsection{Latency, DB Requests, and Hit Rate}
% \label{sec:eval:metrics}

% We sweep the prefetch limit across a workload-appropriate range and aggregate each metric by the configurations plotted for that figure. Figure~\ref{fig:sweep-all} reports median end-to-end request latency for TTG-disabled, MSP, and TTG-enabled configurations. In the refreshed latency panels, the plot uses one-sided p50-to-p90 whiskers, baseline reference lines for no-prefetch and MSP execution, and a TTG-enabled curve across positive prefetch limits. Figure~\ref{fig:random-paired-average-times} reports median walker time for the random-paired stream experiment, and Table~\ref{tab:selep-random-paired-hit-rate} reports SeLeP's block-cache hit-rate diagnostics on the same stream. Figure~\ref{fig:db-requests-all} is the only Oracle comparison: it reports request-time backing-store pressure as profile-derived PostgreSQL backend calls for None, MSP, Oracle, and TTG. Figure~\ref{fig:ttg-cache-plan-quality} reports TTG runtime-cache hit rate across the prefetch-limit sweep.

\subsection{Prefetch-Limit Sweeps}
\label{sec:eval:metrics}

\textbf{Protocol.} Before each trial, the harness restarts the Jac server on Machine 1, creating a fresh application process, and restores the application's PostgreSQL dump on Machine 2. It then issues a single walker request; measured requests do not overlap. We report the median across $N=10$ trials per configuration for LinkedList, Jacord, and JDrive, and $N=30$ trials per configuration for LittleX. The server uses CPython~3.14's free-threaded build, with cyclic garbage collection disabled through \texttt{JAC\_\allowbreak DISABLE\_\allowbreak GC=1}.

\begin{figure*}[t]
\centering
\includegraphics[width=\linewidth]{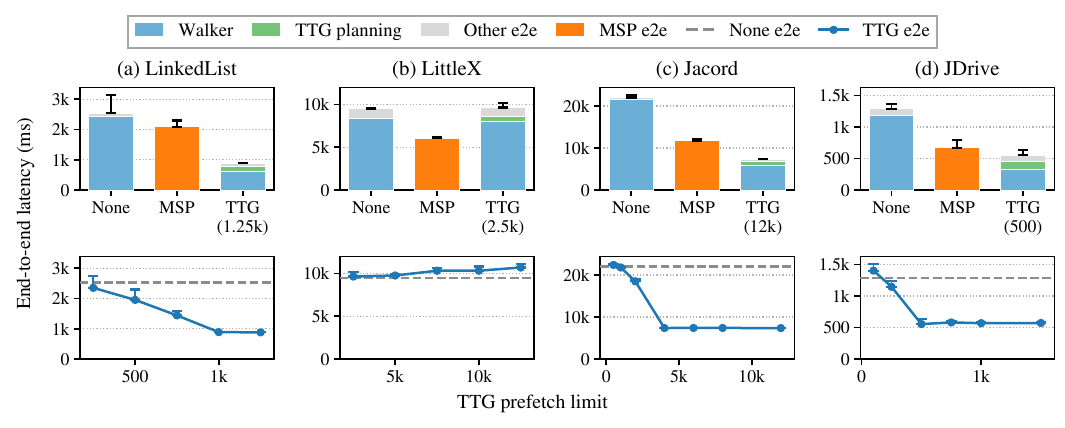}
\caption{End-to-end request latency. Top: None, MSP, and TTG (lowest median latency, limit in parentheses). Stacks show walker, planning, and remaining time; MSP shows total time. Bottom: TTG across all measured positive limits and the None median (dashed). Bars and TTG points show medians, with whiskers to p90. Each column shares a vertical scale.}
\Description{Four columns correspond to LinkedList, LittleX, Jacord, and JDrive. The top row compares no prefetching, manually staged prefetching, and the TTG configuration with the lowest median end-to-end latency using bars. The bottom row connects median TTG end-to-end latency at every measured positive prefetch limit, with a dashed horizontal line for the no-prefetch median. Bar and point whiskers extend from the median to p90. Each column uses the same vertical scale for both rows. One shared legend identifies the bar components and the two line styles.}
\label{fig:sweep-all}
\end{figure*}

For each application, we vary the TTG's prefetch limit and compare its end-to-end request latency with the stock runtime and the MSP baseline. Figure~\ref{fig:sweep-all} compares None, MSP, and TTG above the full prefetch-limit sweep. We also examine backing-store request counts and runtime-cache hit rates to characterize how prefetching changes storage-access activity and data availability when the walker requests an object.

\begin{figure*}[t]
\centering
\includegraphics[width=\linewidth]{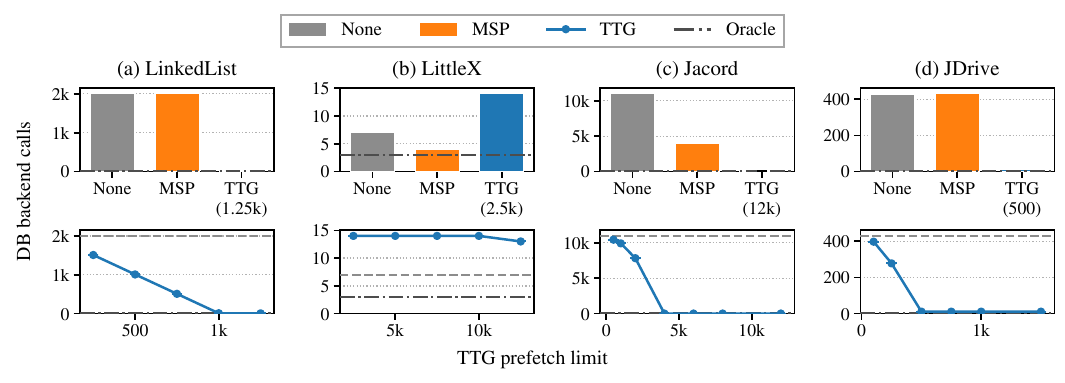}
\caption{Backing-store request pressure. Layout and TTG configurations follow Figure~\ref{fig:sweep-all}. Bars and points show median profiled PostgreSQL backend calls (whiskers: p90). Reference lines: None (dashed), Oracle (dash-dotted).}
\Description{Four columns show LinkedList, LittleX, Jacord, and JDrive. Upper bars compare no prefetching, manually staged prefetching, and the TTG configuration with the lowest median end-to-end latency. Lower lines connect median PostgreSQL backend-call counts at all measured positive TTG limits against a dashed no-prefetch median. A dash-dotted Oracle median appears in both rows. Bar and TTG point whiskers extend to p90. Each column shares a vertical scale, and all panels share one legend.}
\label{fig:db-requests-all}
\end{figure*}

\begin{figure*}[t]
\centering
\includegraphics[width=\linewidth]{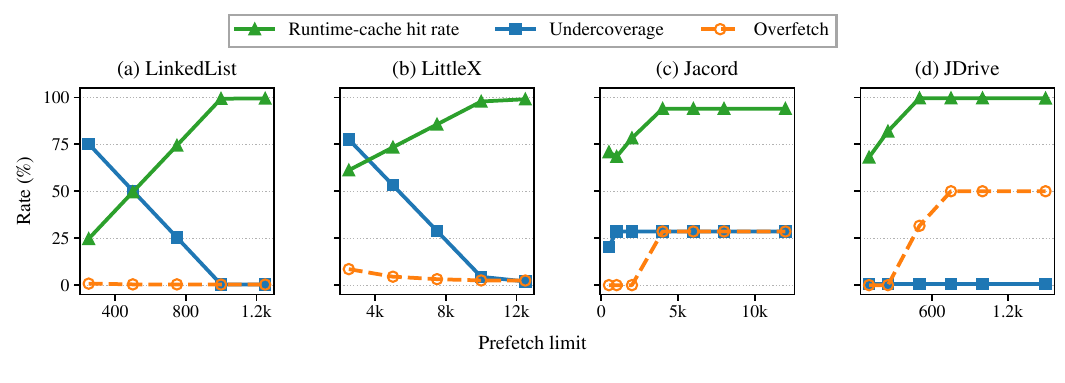}
\caption{Runtime-cache hit rate and cache-derived diagnostics under TTG.}
\Description{Four panels show LinkedList, LittleX, Jacord, and JDrive. Each panel plots runtime-cache hit rate with green triangles, undercoverage with blue squares, and overfetch with orange open circles across the TTG prefetch-limit sweep. The undercoverage and overfetch curves are cache-derived diagnostics computed from distinct accessed UUIDs, distinct cache-hit UUIDs, and reported plan size.}
\label{fig:ttg-cache-plan-quality}
\end{figure*}

\textbf{LinkedList.} Figure~\ref{fig:sweep-all}(a) shows that TTG reduces median end-to-end latency from \obsLinkedBaselineMs~ms to \obsLinkedBestMs~ms at prefetch limit \obsLinkedBestLimit{}, achieving a $\speedupLinked\times$ speedup over the stock runtime. At this configuration, profiled PostgreSQL backend calls decrease from \obsLinkedDbBaseline{} to \obsLinkedDbBest{} (Figure~\ref{fig:db-requests-all}(a)), while the runtime-cache hit rate increases from \obsLinkedLoneBaselineHit\% to \obsLinkedLoneBestHit\% (Figure~\ref{fig:ttg-cache-plan-quality}(a)). The walker's repeated \texttt{Next} visits expose a continuation that the Builder can expand into a chain prefix before execution, allowing successive node accesses to reuse resolved topology and prefetched data.

\textbf{LittleX.} TTG provides no end-to-end speedup in this sweep. Even its lowest median latency is \obsLittleXBestMs~ms at prefetch limit \obsLittleXBestLimit{}, compared with \obsLittleXBaselineMs~ms for the stock runtime: a $\speedupLittleX\times$ speedup, or a measured slowdown (Figure~\ref{fig:sweep-all}(b)). Increasing the limit to \obsLittleXHighHitLimit{} raises the run\-time-cache hit rate to \obsLittleXHighHit\% (Figure~\ref{fig:ttg-cache-plan-quality}(b)). Nevertheless, end-to-end latency increases to \obsLittleXHighHitMs~ms ($\obsLittleXHighHitSpeedup\times$). At this budget, median synchronous planning time is \obsLittleXHighHitPlanningMs~ms, while median walker time decreases by only \obsLittleXHighHitWalkerReductionMs~ms. The stock runtime already serves the request with \obsLittleXDbBaseline{} backend calls; TTG increases this count to \obsLittleXDbBest{} at its fastest configuration and \obsLittleXHighHitDb{} at the largest budget (Figure~\ref{fig:db-requests-all}(b)). LittleX therefore demonstrates a systems boundary: a high runtime-cache hit rate is insufficient when planning overhead exceeds the demand latency avoided.

\textbf{Jacord.} Figure~\ref{fig:sweep-all}(c) shows that TTG reduces the median end-to-end latency of \texttt{load\_channel} from \obsJacordBaselineMs~ms to \allowbreak\obsJacordBestMs~ms at prefetch limit \obsJacordBestLimit{}. The resulting $\speedupJacord\times$ speedup is the largest among the four applications. Profiled PostgreSQL backend calls decrease from \obsJacordDbBaseline{} to \obsJacordDbBest{} (Figure~\ref{fig:db-requests-all}(c)). This workload recursively follows incoming \texttt{ReplyTo} edges and reads author associations for each message. The traversal and ability triggers allow the Builder to compose these accesses across successive reply levels, identifying candidate messages and authors before demand execution reaches them. At smaller limits of \obsJacordSubBaselineLimitA{} and \obsJacordSubBaselineLimitB{}, speedups remain near baseline at $\obsJacordSubBaselineSpeedupA\times$ and $\obsJacordSubBaselineSpeedupB\times$, respectively.

\textbf{JDrive.} Figure~\ref{fig:sweep-all}(d) shows that TTG reduces median end-to-end latency from \obsJDriveBaselineMs~ms to \obsJDriveBestMs~ms at prefetch limit \obsJDriveBestLimit{}, achieving a $\speedupJDrive\times$ speedup. Profiled PostgreSQL backend calls decrease from \obsJDriveDbBaseline{} to \obsJDriveDbBest{} (Figure~\ref{fig:db-requests-all}(d)). The \texttt{Visible\allowbreak FolderTree} walker follows \texttt{Contains} edges to child folders whose \texttt{trashed} flag is false, pruning subtrees rooted at trashed children. The Builder expands the typed traversal without evaluating this predicate, so its candidate plan can include folders the request never reads. The measured speedup shows that useful prefetching remains possible with this conservative plan; Figure~\ref{fig:ttg-cache-plan-quality}(d) reports the corresponding runtime-cache hit rate and cache-derived diagnostics. At limit \obsJDriveSubBaselineLimitA{}, TTG yields $\obsJDriveSubBaselineSpeedupA\times$, increasing to $\obsJDriveSubBaselineSpeedupB\times$ at limit \obsJDriveSubBaselineLimitB{}.

\begin{figure*}[t]
\centering
\includegraphics[width=\linewidth]{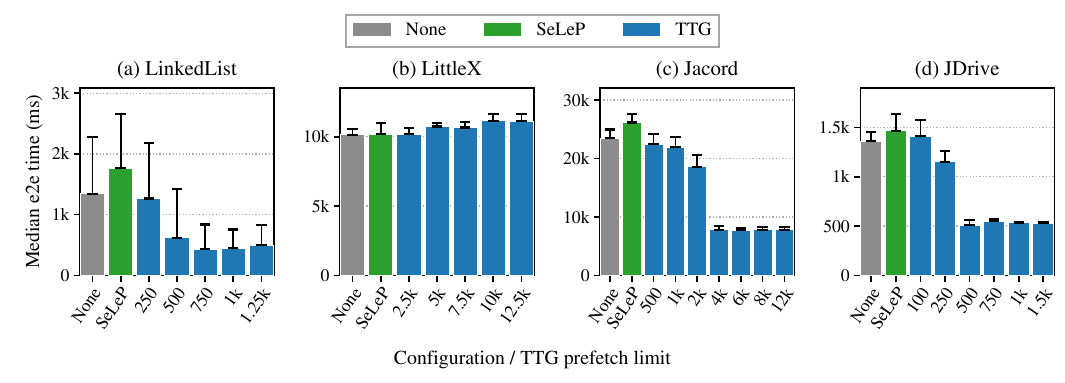}
\caption{End-to-end request latency on randomized request streams. After SeLeP training, the no-prefetch, SeLeP, and TTG configurations execute the same sequence of 20 randomly selected requests per application. Bars show median client-measured latency, including synchronous TTG planning when enabled; whiskers extend to p90.}
\Description{Four panels show LinkedList, LittleX, Jacord, and JDrive. Gray bars denote no prefetching, green bars SeLeP, and blue bars TTG across all measured prefetch limits. Heights show median end-to-end request latency and whiskers extend to the 90th percentile.}
\label{fig:random-paired-average-times}
\end{figure*}

\textbf{Explanation.} TTG's end-to-end benefit depends on how much demand-time storage waiting it avoids relative to the costs of planning and prefetching. Synchronous plan construction lies on the request's critical path, while background reads consume runtime and database resources. Increasing the prefetch limit can improve coverage, but can also increase topology-resolution and object-loading work. Moreover, inclusion in the plan does not guarantee a cache hit: prefetching may still be in progress when the walker requests an object. LinkedList, Jacord, and JDrive exhibit both fewer backing-store calls and lower end-to-end latency. LittleX demonstrates that even a high runtime-cache hit rate need not produce a latency benefit when the baseline already serves the request with few backing-store calls.

\textbf{Cache-derived diagnostics.} Figure~\ref{fig:ttg-cache-plan-quality} combines runtime-cache hit rate with diagnostics computed from access logs and plan size. Let $a$ count distinct nonzero UUIDs in the request's access log, $h$ count those with at least one runtime-cache hit, and $p$ be the reported plan size. The curves labeled undercoverage and overfetch plot $(a-h)/a$ and $\max(p-h,0)/p$, respectively, as percentages. Cache hits on both prefetched and previously demand-loaded objects contribute to $h$. For LinkedList and LittleX, increasing the budget reduces undercoverage with little overfetch. JDrive's cache-derived undercoverage stays near \obsJDriveUndercoverage\%, while its cache-de\-rived overfetch reaches \obsJDriveOverfetch\% at the largest measured budget.

Jacord's plateau reflects prefetch timeliness. The walker reaches its initial \texttt{visit} before the corresponding prefetched objects are available and loads the first message batch on demand. Objects without a subsequent cache hit remain outside $h$. Once the plan size saturates at $p=a$, the two cache-derived diagnostics coincide; increasing the bound does not give this initial visit additional prefetch lead time.

\textbf{Takeaway.} OSP's explicit traversal continuation enables the TTG Builder to construct request-specific plans spanning multiple walker steps. In our evaluation, these plans yield end-to-end speedups over the stock runtime in three of the four applications, reaching up to $\headlineSpeedup\times$. Their utility depends on both candidate selection and execution costs. Value-dependent pruning can leave planned objects unused, as in JDrive, while high runtime-cache hit rates can coexist with no latency improvement, as in LittleX. The prefetch limit bounds the candidate set; increasing it can improve coverage but does not guarantee lower latency.

\subsection{Comparison with SeLeP}
\label{sec:eval:selep}

\textbf{Protocol.} Before measurement, we train SeLeP using SQL traces collected from \selepLinkedTrainRequests{}, \selepLittleXTrainRequests{}, \selepJacordTrainRequests{}, and \selepJDriveTrainRequests{} application requests for LinkedList, LittleX, Jacord, and JDrive, respectively. After filtering, these traces contain \selepLinkedTrainSqlEvents{}, \selepLittleXTrainSqlEvents{}, \selepJacordTrainSqlEvents{}, and \selepJDriveTrainSqlEvents{} SQL events, respectively. For each application, we then execute the no-prefetch, SeLeP, and TTG configurations on the same sequence of 20 randomly selected requests. Jac and PostgreSQL remain running throughout each request stream. SeLeP observes the SQL trace generated across the full stream. The no-prefetch and TTG configurations follow the same execution protocol. We report the median \texttt{e2e\_ms} across the stream with p50--p90 variation bars. This client-side measurement includes synchronous TTG planning when enabled. To separate SeLeP's prediction quality from request latency, we also evaluate its block predictions using trace-driven block-cache simulation.

\textbf{End-to-end latency.} Figure~\ref{fig:random-paired-average-times} compares end-to-end latency on these request streams. Under the evaluated configuration, SeLeP does not reduce median request latency relative to the no-prefetch baseline. With sufficient prefetch budgets, TTG reduces median latency on LinkedList, Jacord, and JDrive. LittleX remains a boundary case: neither approach improves its median end-to-end latency.

\textbf{Block-cache simulation.} We compute the block-cache hit rates in Table~\ref{tab:selep-random-paired-hit-rate} using SeLeP's original hit-rate evaluation method~\cite{selep}, reusing its LRU-cache implementation. The simulation processes the measured demand-block access stream with and without SeLeP prefetches. Hit rate is the fraction of demand block accesses found in the simulated cache. SeLeP eliminates most misses in the LinkedList and Jacord simulations, despite the absence of an end-to-end latency reduction. On LittleX and JDrive, miss reduction is limited and many prefetched blocks are unused. These diagnostics distinguish useful block predictions from their effect on request execution. Warming database blocks can reduce storage I/O while leaving dependent application--database round trips intact. TTG instead makes planned object data and resolved topology available within the Jac runtime. For both approaches, the performance benefit depends on whether avoided demand-time waiting outweighs the cost of planning and prefetching.

\input{figures/selep-random-paired-hit-rates.tex}

\section{Related Work}
\label{sec:rw}

Existing prefetchers occupy several points in the design space: programmer annotations, profile-guided application-layer inference, compile-time rewriting, reactive batching, and history-based prediction. 
% Table~\ref{tab:design-space} summarises what each family can use when it chooses work to fetch: the application or walker source, the current live graph, and the current request. TTG's contribution is that the runtime layer can combine all three before the walker executes, without programmer effort.

% \begin{table}[t]
% \centering
% \caption{What each prefetcher family can use when choosing prefetched work. \seesauto: seen automatically; \seesprog: seen only via a programmer annotation; \notseen: not seen at plan time. Profiled, prefix, and history denote weaker signals from previous executions, the already-reached request prefix, and previous requests.}
% \label{tab:design-space}
% \scriptsize
% \begin{tabular}{@{}p{0.42\columnwidth}ccc@{}}
% \toprule
%                                                  & Source     & Live graph & Request    \\
% \midrule
% Annotations \textit{\scriptsize(Django, DataLoader, Rails)} & \seesprog & \seesprog & \seesprog \\
% Profile-guided \textit{\scriptsize(AutoFetch)}              & profiled  & \notseen  & \notseen  \\
% Compile-time rewriting \textit{\scriptsize(DBridge)}        & \seesauto & \notseen  & \notseen  \\
% Reactive batching \textit{\scriptsize(Sloth)}               & \seesauto & \notseen  & prefix    \\
% History prediction \textit{\scriptsize(Pred-Cache, SeLeP)}  & \notseen  & \notseen  & history   \\
% TTG Builder                                                 & \seesauto & \seesauto & \seesauto \\
% \bottomrule
% \end{tabular}
% \end{table}

{\emergencystretch=1em
\textbf{Programmer-directed mechanisms} expose access information through hints and programming interfaces~\cite{informed-prefetching,django-select-related,rails-includes,graphql-dataloader,minnow}. Foreactor~\cite{foreactor} uses explicit I/O dependency graphs, while PULSE~\cite{pulse} uses developer-defined iterators to execute pointer traversals near disaggregated memory. TTG derives a bounded candidate plan from OSP traversal semantics and the live graph while keeping walker execution client-side.\par}

\textbf{Profile-guided prefetchers} infer prefetches from earlier executions. AutoFetch~\cite{autofetch} profiles ORM traversals; prediction quality can decrease when request traversals differ from prior profiles.

\textbf{Compile-time rewriting} uses program analysis to batch database calls~\cite{batched-bindings,dbridge,reformulator}; DBridge also introduces asynchronous query prefetching~\cite{dbridge-prefetch}. Other work derives database queries or SQL expressions from program code~\cite{query-extraction,interprocedural-query-extraction,qbs,clis}, with interprocedural query extraction composing traversal summaries at runtime~\cite{interprocedural-query-extraction}. Compiler-assisted prefetching also targets dependent processor-memory accesses~\cite{event-triggered}. TTG derives a bounded plan from OSP traversal-continuation rules and the live graph at spawn time.

\textbf{Reactive batching} groups database operations during execution~\cite{transactionmerger}. Sloth~\cite{sloth} defers query execution and dependent computations through extended lazy evaluation, issuing accumulated queries together when a result is required. TTG constructs a bounded candidate plan before walker execution.

\textbf{Code motion across the application--storage boundary} includes application partitioning~\cite{pyxis} and compilation of imperative control flow with SQL-based computations into read-only SQL~\cite{flummi}. WeBridge~\cite{webridge} uses concolic execution on hot paths to synthesize stored procedures containing queries and supported computations of their parameters and path conditions. TTG plans candidate reads while keeping walker execution client-side, without requiring stored procedures. Its conservative treatment of value-dependent control can introduce unused candidates, whose number is bounded by the plan budget.

{\emergencystretch=1em
\textbf{Workload-driven prefetching and caching} use observed query and data-access patterns~\cite{compression-prefetching,predcache,pre-buffer,grasp}. SeLeP~\cite{selep} uses block-content encodings to predict future partition accesses. Pythia~\cite{pythia} predicts non-sequential page accesses from the current query plan using models trained on query access traces. TTG constructs a bounded object-level plan from explicit traversal rules and the live graph rooted at the current request.\par}

\section{Discussion}
\label{sec:discussion}

\textbf{When planning pays off.} TTG improves latency when avoided demand stalls outweigh planning and prefetching costs. LinkedList, Jacord, and JDrive benefit from planning dependent accesses ahead of execution, even with JDrive's value-dependent overfetch. LittleX shows that a high run\-time-cache hit rate is insufficient when the stock runtime already uses few backing-store calls and synchronous planning exceeds the demand latency avoided (\S\ref{sec:eval:metrics}).

\textbf{Comparison with MSP.} MSP already combines manually specified association paths, asynchronous submission, dependency chaining, and compatible batching. TTG uses explicit continuation rules to identify candidate objects across future walker steps before execution, exposing opportunities to batch reads across those steps and reuse resolved topology. With sufficient prefetch budgets, TTG achieves lower median end-to-end latency than MSP on LinkedList, Jacord, and JDrive, while MSP is faster on LittleX (Figure~\ref{fig:sweep-all}). MSP combines LittleX's feed paths into a single topology query and batches their object loads, reducing round trips without TTG plan construction. LittleX already uses few batched demand queries, leaving limited opportunity to offset TTG's planning and prefetching costs; its multi-hop feed query also remains on the demand path in the evaluated TTG implementation. These results characterize the benefits and costs of continuation-based planning across walker steps relative to manually optimized staged execution.

\textbf{Semantic boundary.} TTG plans structural dependencies through typed associations and continuation rules, while conservatively approximating value-dependent control. In JDrive, the Builder expands \texttt{Contains} paths without evaluating \texttt{trashed}, so it can include descendants of folders that the walker prunes. The prefetch limit $L$ bounds the number of unused candidates, not their fraction of the plan.

\textbf{Design trade-offs.} The static analyzer overapproximates in one direction that has a real cost: when a walker's visit type is a runtime variable, the analyzer falls back to the schema-wide edge set (\S\ref{sec:impl:analyzer}), which can inflate the plan by the number of declared edge classes. Our benchmarks all use literal edge types, so we do not measure this cost; a deployment with heavy runtime-variable use would see it. Conversely, the TTG-generated GTI (\S\ref{sec:impl:builder}) adds session-local memory proportional to the topology rows returned by the planning query; it is not a standing root-persisted index in the evaluated implementation.

The current implementation uses a fixed plan bound and our evaluation focuses on read-dominated, sequential requests. Concurrent workloads may introduce contention in the prefetch workers and shared backing store, while concurrent graph updates require additional cache-consistency mechanisms. We leave adaptive planning and concurrent execution to future work.

\section{Conclusion}
\label{sec:conclusion}

{\emergencystretch=1em
This work identifies a programming-model contract for request-specific read planning: inspectable access paths paired with explicit traversal continuation. OSP provides one instantiation, and TTG composes its access and continuation rules over the live graph at request spawn into bounded candidate plans spanning future traversal steps. Our Jac/PostgreSQL evaluation across four workloads achieves up to $\headlineSpeedup\times$ end-to-end speedup over the stock runtime and characterizes trade-offs relative to manually staged prefetching. LittleX shows that planning overhead can exceed the demand latency avoided, while JDrive shows that value-dependent control can introduce bounded overfetch. Together, these results demonstrate the practical value and workload-dependent limits of exposing traversal continuation as an interface for request-specific storage planning.\par}

\section*{Acknowledgments}

We used OpenAI Codex to assist with manuscript drafting and revision, literature search, code development, and figure preparation. The authors take full responsibility for the content, results, and citations.

\bibliographystyle{ACM-Reference-Format}
\bibliography{references}

\end{document}

%% file: evaluation_macros.tex
\newcommand{\speedupLinked}{2.87}
\newcommand{\speedupLittleX}{0.98}
\newcommand{\speedupJacord}{2.99}
\newcommand{\speedupJDrive}{2.32}
\newcommand{\headlineSpeedup}{\speedupJacord}

\newcommand{\obsLinkedBaselineMs}{2529}
\newcommand{\obsLinkedBestMs}{883}
\newcommand{\obsLinkedBestLimit}{1250}

\newcommand{\obsLinkedDbBaseline}{2000}
\newcommand{\obsLinkedDbBest}{7}
\newcommand{\obsLinkedLoneBaselineHit}{0.0}
\newcommand{\obsLinkedLoneBestHit}{99.4}
\newcommand{\obsLittleXBaselineMs}{9491}
\newcommand{\obsLittleXBestMs}{9638}
\newcommand{\obsLittleXBestLimit}{2500}

\newcommand{\obsLittleXDbBaseline}{7}
\newcommand{\obsLittleXDbBest}{14}
\newcommand{\obsLittleXHighHitLimit}{12500}
\newcommand{\obsLittleXHighHit}{99.0}
\newcommand{\obsLittleXHighHitDb}{13}
\newcommand{\obsLittleXHighHitMs}{10688}
\newcommand{\obsLittleXHighHitSpeedup}{0.89}
\newcommand{\obsLittleXHighHitPlanningMs}{1393}
\newcommand{\obsLittleXHighHitWalkerReductionMs}{341}

\newcommand{\obsJacordBaselineMs}{21982}
\newcommand{\obsJacordBestMs}{7362}
\newcommand{\obsJacordBestLimit}{12000}

\newcommand{\obsJacordDbBaseline}{10982}
\newcommand{\obsJacordDbBest}{12}
\newcommand{\obsJacordSubBaselineLimitA}{500}
\newcommand{\obsJacordSubBaselineSpeedupA}{0.98}

\newcommand{\obsJacordSubBaselineLimitB}{1000}
\newcommand{\obsJacordSubBaselineSpeedupB}{1.01}

\newcommand{\obsJDriveBaselineMs}{1288}
\newcommand{\obsJDriveBestMs}{555}
\newcommand{\obsJDriveBestLimit}{500}
\newcommand{\obsJDriveDbBaseline}{429}
\newcommand{\obsJDriveDbBest}{11}
\newcommand{\obsJDriveSubBaselineLimitA}{100}
\newcommand{\obsJDriveSubBaselineSpeedupA}{0.92}
\newcommand{\obsJDriveSubBaselineLimitB}{250}
\newcommand{\obsJDriveSubBaselineSpeedupB}{1.12}

\newcommand{\obsJDriveUndercoverage}{0.6}
\newcommand{\obsJDriveOverfetch}{50}

\newcommand{\selepLinkedTrainRequests}{100}
\newcommand{\selepLinkedTrainSqlEvents}{101,901}
\newcommand{\selepLittleXTrainRequests}{20}
\newcommand{\selepLittleXTrainSqlEvents}{181}
\newcommand{\selepJacordTrainRequests}{20}
\newcommand{\selepJacordTrainSqlEvents}{101,318}
\newcommand{\selepJDriveTrainRequests}{30}
\newcommand{\selepJDriveTrainSqlEvents}{5,281}

%% file: figures/plan_example.tikz
% Side-by-side panels in one column, with unscaled 8--9pt text.
\begin{tikzpicture}[
  x=1cm, y=1cm,
  font=\sffamily\fontsize{8}{9.2}\selectfont,
  every node/.style={inner sep=0pt, outer sep=0pt},
  graphnode/.style={circle, align=center, minimum size=0.55cm,
                    font=\sffamily\fontsize{9}{10.5}\selectfont},
  commit/.style   ={graphnode, draw=blue!55, fill=blue!10, thick},
  addnew/.style   ={graphnode, draw=green!55!black, fill=green!15, very thick},
  skip/.style     ={graphnode, draw=black!55, fill=white, dashed,
                    text=black!70},
  faded/.style    ={graphnode, draw=black!20, fill=black!3, thin,
                    text=black!30},
  neutral/.style  ={graphnode, draw=black!45, fill=black!6, thin},
  efollow/.style  ={-{Latex[length=1.4mm]}, draw=blue!55,   line width=0.55pt},
  epost/.style    ={-{Latex[length=1.4mm]}, draw=green!55!black,
                    line width=0.55pt},
  edim/.style     ={-{Latex[length=1.4mm]}, draw=black!18,  line width=0.4pt},
  edgelab/.style  ={font=\sffamily\fontsize{8}{9.2}\selectfont\itshape,
                    inner xsep=1.5pt, inner ysep=0.6pt},
  levellab/.style ={font=\sffamily\fontsize{8}{9.2}\selectfont\bfseries},
  planbox/.style  ={draw=black!35, fill=black!4, rounded corners=1.5pt,
                    inner xsep=6pt, inner ysep=4pt, align=center,
                    font=\sffamily\fontsize{8}{9.2}\selectfont},
  paneltitle/.style={font=\sffamily\fontsize{9}{10.5}\selectfont\bfseries,
                    align=center},
  legendtext/.style={font=\sffamily\fontsize{8}{9.2}\selectfont,
                    align=left},
]

\def\panelspan{4.4}

% Shared top-down topology, including the disconnected c/t5 pair.
\newcommand{\layout}{%
  \coordinate (u0) at (1.55, 2.3);
  \coordinate (a)  at (0.7, 1.15);
  \coordinate (b)  at (2.3, 1.15);
  \coordinate (c)  at (3.5, 2.3);
  \coordinate (t1) at (0.25, 0);
  \coordinate (t2) at (1.05, 0);
  \coordinate (t3) at (1.9, 0);
  \coordinate (t4) at (2.7, 0);
  \coordinate (t5) at (3.5, 1.15);
}

% Edge list drawing macro.  Three style tokens let each panel
% recolour: #1 = Follows edges, #2 = Posted edges, #3 = the
% (c) --> (t5) Posted edge (dimmed in panel (b) since c is not
% reachable via the walker's typed chain from u0).
\newcommand{\drawgraph}[3]{%
  \draw[#1] (v-u0) -- (v-a);
  \draw[#1] (v-u0) -- (v-b);
  \draw[#2] (v-a)  -- (v-t1);
  \draw[#2] (v-a)  -- (v-t2);
  \draw[#2] (v-b)  -- (v-t3);
  \draw[#2] (v-b)  -- (v-t4);
  \draw[#3] (v-c)  -- (v-t5);
}

\begin{scope}[shift={(0, 0)}]
  \node[paneltitle] at (1.9, 3.1) {(a) OSP graph};

  \layout

  \node[neutral] (v-u0) at (u0) {$u_0$};
  \node[neutral] (v-a)  at (a)  {$a$};
  \node[neutral] (v-b)  at (b)  {$b$};
  \node[neutral] (v-c)  at (c)  {$c$};
  \node[neutral] (v-t1) at (t1) {$t_1$};
  \node[neutral] (v-t2) at (t2) {$t_2$};
  \node[neutral] (v-t3) at (t3) {$t_3$};
  \node[neutral] (v-t4) at (t4) {$t_4$};
  \node[neutral] (v-t5) at (t5) {$t_5$};
  \drawgraph{efollow}{epost}{epost}
\end{scope}

\begin{scope}[shift={(\panelspan, 0)}]
  \node[paneltitle] at (1.9, 3.1) {(b) TTG plan};

  \layout

  \node[commit] (v-u0) at (u0) {$u_0$};
  \node[skip]   (v-a)  at (a)  {$a$};
  \node[skip]   (v-b)  at (b)  {$b$};
  \node[faded]  (v-c)  at (c)  {$c$};
  \node[addnew] (v-t1) at (t1) {$t_1$};
  \node[addnew] (v-t2) at (t2) {$t_2$};
  \node[addnew] (v-t3) at (t3) {$t_3$};
  \node[addnew] (v-t4) at (t4) {$t_4$};
  \node[faded]  (v-t5) at (t5) {$t_5$};
  \drawgraph{efollow}{epost}{edim}

  \node[levellab, text=blue!60, anchor=west] at (0, 2.3) {Level 0};
  \node[levellab, text=green!45!black] at (1.5, -0.5) {Level 1};
\end{scope}

\node[planbox, text width=7.75cm] at (4.075, -1.04)
  {TTG prefetch list (BFS order): $[\,u_0,\ t_1,\ t_2,\ t_3,\ t_4\,]$};

% Shared type and edge keys replace repeated labels inside both panels.
\node[legendtext, anchor=west] at (0, -1.68) {User: $u_0,a,b,c$};
\node[legendtext, anchor=west] at (2.12, -1.68) {Tweet: $t_1,\ldots,t_5$};
\draw[efollow] (4.55, -1.68) -- (5.03, -1.68);
\node[edgelab, anchor=west] at (5.15, -1.68) {Follows};
\draw[epost] (6.62, -1.68) -- (7.10, -1.68);
\node[edgelab, anchor=west] at (7.22, -1.68) {Posted};

\begin{scope}[shift={(0, -2.25)}]
  \def\sw{0.28}

  \node[commit, minimum size=\sw cm, inner sep=0pt]
        (l1) at (0.1, 0) {};
  \node[right=3pt of l1, legendtext] {Spawn};

  \node[addnew, minimum size=\sw cm, inner sep=0pt]
        (l2) at (1.85, 0) {};
  \node[right=3pt of l2, legendtext] {Planned};

  \node[skip, minimum size=\sw cm, inner sep=0pt]
        (l3) at (3.85, 0) {};
  \node[right=3pt of l3, legendtext] {Skip-eligible};

  \node[faded, minimum size=\sw cm, inner sep=0pt]
        (l4) at (6.25, 0) {};
  \node[right=3pt of l4, legendtext] {Not reached};
\end{scope}
\end{tikzpicture}

%% file: figures/workload_graphs.tikz
% OSP graph shapes for the four evaluation workloads (Sec 4.1).
%
% Single-row layout. Each panel shows a small concrete instance of
% the workload's graph so the reader can see the node types, edge
% types, and characteristic shape at a glance.
%
%   (a) LinkedList : Item -Next-> Item -Next-> ... chain
%   (b) LittleX    : User -Follows-> User -Posted-> Tweet
%   (c) Jacord     : Channel -contains-> Message -reply-> Message*
%   (d) JDrive     : Folder -Contains-> Folder*, value-pruned
%
% Node/edge styling mirrors builder_progression.tikz and
% plan_example.tikz so all Sec 3.4/4.1 figures share one visual
% language.
% Compress coordinates rather than scaling text: labels remain 9pt.
\begin{tikzpicture}[
  x=0.58cm, y=0.58cm,
  font=\sffamily\small,
  every node/.style={inner sep=0pt, outer sep=0pt},
  % ---- node styles ----
  gnode/.style   ={draw=black!45,        fill=black!6,      thin,
                   circle, align=center,
                   minimum size=0.45cm, inner sep=0pt,
                   font=\sffamily\small},
  spawn/.style   ={draw=blue!55,         fill=blue!10,      thick,
                   circle, align=center,
                   minimum size=0.45cm, inner sep=0pt,
                   font=\sffamily\small},
  pruned/.style  ={draw=black!30,        fill=black!3,      dashed,
                   text=black!65, thin, circle, align=center,
                   minimum size=0.45cm, inner sep=0pt,
                   font=\sffamily\small},
  tweet/.style   ={gnode, rectangle, rounded corners=1pt,
                   minimum width=0.50cm, minimum height=0.42cm},
  gedge/.style   ={-{Latex[length=1.4mm]}, draw=black!55, line width=0.55pt},
  epruned/.style ={-{Latex[length=1.4mm]}, draw=black!35, dashed, line width=0.55pt},
  % LittleX typed edges (match plan_example.tikz palette).
  efollow/.style ={-{Latex[length=1.4mm]}, draw=blue!55,   line width=0.55pt},
  epost/.style   ={-{Latex[length=1.4mm]}, draw=green!55!black,
                                                          line width=0.55pt},
  edgelab/.style ={font=\sffamily\small\itshape,
                   inner xsep=1.5pt, inner ysep=0.6pt},
  paneltitle/.style={font=\sffamily\small\bfseries, align=center},
  panelnote/.style={font=\sffamily\small, text=black!75,
                    text width=4.05cm, align=center, anchor=north},
  paneldots/.style ={font=\sffamily\normalsize, text=black!55},
]

\def\panelspan{7.6}

% ===============================================================
% Panel (a): LinkedList -- deep chain along a single edge type
% ===============================================================
\begin{scope}[shift={(0, 0)}]
  \node[paneltitle] at (2.8, 5.6)
    {(a) \emph{LinkedList}};

  \node[spawn] (n0) at (0.4, 2.4) {$n_0$};
  \node[gnode] (n1) at (1.7, 2.4) {$n_1$};
  \node[gnode] (n2) at (3.0, 2.4) {$n_2$};
  \node[gnode] (n3) at (4.3, 2.4) {$n_3$};
  \node[paneldots] (dots) at (5.5, 2.4) {$\cdots$};

  \draw[gedge] (n0) -- (n1);
  \draw[gedge] (n1) -- (n2);
  \draw[gedge] (n2) -- (n3);
  \draw[gedge] (n3) -- (dots);

  \node[panelnote] at (2.8, -0.5)
    {\texttt{Item} nodes; \texttt{Next} edges\\
     \texttt{Traverse} follows the chain};
\end{scope}

% ===============================================================
% Panel (b): LittleX -- fan-out over two typed hops
% ===============================================================
\begin{scope}[shift={(\panelspan, 0)}]
  \node[paneltitle] at (2.8, 5.6)
    {(b) \emph{LittleX}};

  \node[spawn] (u0) at (2.0, 3.4) {$u_0$};
  \node[gnode] (ua) at (3.6, 3.4) {$a$};
  \node[gnode] (ub) at (3.6, 1.6) {$b$};
  \node[gnode] (uc) at (2.0, 1.6) {$c$};

  \node[tweet] (t0) at (0.5, 4.5) {$t_0$};
  \node[tweet] (ta) at (5.1, 4.5) {$t_a$};
  \node[tweet] (tb) at (5.1, 0.5) {$t_b$};
  \node[tweet] (tc) at (0.5, 0.5) {$t_c$};

  \draw[efollow] (u0) -- (ua);
  \draw[efollow] (ua) -- (ub);
  \draw[efollow] (ub) -- (uc);
  \draw[efollow] (uc) -- (u0);

  \draw[epost] (u0) -- (t0);
  \draw[epost] (ua) -- (ta);
  \draw[epost] (ub) -- (tb);
  \draw[epost] (uc) -- (tc);

  \node[panelnote] at (2.8, -0.5)
    {Circles: User; boxes: Tweet\\
     \textcolor{blue!55!black}{\texttt{Follows}} then
     \textcolor{green!55!black}{\texttt{Posted}}};
\end{scope}

% ===============================================================
% Panel (c): Jacord -- reply tree beneath channel messages
% ===============================================================
\begin{scope}[shift={(2*\panelspan, 0)}]
  \node[paneltitle] at (2.8, 5.6)
    {(c) \emph{Jacord}};

  \node[spawn] (ch) at (0.4, 2.4) {\textsc{Ch}};
  \node[gnode] (m1) at (2.4, 3.8) {$m_1$};
  \node[gnode] (m2) at (2.4, 1.0) {$m_2$};
  \node[gnode] (r11) at (4.7, 4.4) {$r_{11}$};
  \node[gnode] (r12) at (4.7, 3.1) {$r_{12}$};
  \node[gnode] (r21) at (4.7, 1.0) {$r_{21}$};

  \draw[gedge] (ch) -- (m1);
  \draw[gedge] (ch) -- (m2);
  \draw[gedge] (m1) -- (r11);
  \draw[gedge] (m1) -- (r12);
  \draw[gedge] (m2) -- (r21);

  \node[paneldots] (rdots) at (5.6, 4.4) {$\cdots$};
  \draw[gedge] (r11) -- (rdots);

  \node[panelnote] at (2.8, -0.5)
    {Ch: Channel; $m/r$: Message\\
     \texttt{contains} / recursive \texttt{reply}};
\end{scope}

% ===============================================================
% Panel (d): JDrive -- folder tree with value-pruned branches
% ===============================================================
\begin{scope}[shift={(3*\panelspan, 0)}]
  \node[paneltitle] at (2.8, 5.6)
    {(d) \emph{JDrive}};

  \node[spawn] (root) at (0.5, 2.6) {$f_0$};
  \node[gnode] (fa) at (2.2, 4.0) {$f_a$};
  \node[gnode] (fb) at (2.2, 1.2) {$f_b$};
  \node[gnode] (faa) at (4.2, 4.6) {$f_{aa}$};
  \node[gnode] (fab) at (4.2, 3.3) {$f_{ab}$};
  \node[pruned] (ftrash) at (4.2, 1.2) {$f_t$};
  \node[paneldots] (fdots) at (5.5, 4.6) {$\cdots$};

  \draw[gedge] (root) -- (fa);
  \draw[gedge] (root) -- (fb);
  \draw[gedge] (fa) -- (faa);
  \draw[gedge] (fa) -- (fab);
  \draw[epruned] (fb) -- (ftrash);
  \draw[gedge] (faa) -- (fdots);

  \node[panelnote] at (2.8, -0.5)
    {Folder nodes; \texttt{Contains} edges\\
     \texttt{trashed} prunes branches};
\end{scope}

% ===============================================================
% Shared legend
% ===============================================================
\begin{scope}[shift={(1.7, -2.45)}]
  \def\dx{9.0}
  \node[spawn, minimum size=0.28cm, inner sep=0pt]
        (ls) at (0, 0) {};
  \node[right=3pt of ls, font=\sffamily\small]
        {walker spawn node};

  \node[gnode, minimum size=0.28cm, inner sep=0pt]
        (lg) at (\dx, 0) {};
  \node[right=3pt of lg, font=\sffamily\small]
        {ordinary graph node};

  \node[pruned, minimum size=0.28cm, inner sep=0pt]
        (lp) at (2*\dx, 0) {};
  \node[right=3pt of lp, font=\sffamily\small]
        {value-pruned branch};
\end{scope}
\end{tikzpicture}

%% file: figures/selep-random-paired-hit-rates.tex
\begin{table}[htbp]
\centering
\caption{SeLeP block-cache diagnostics for randomized request streams (\S\ref{sec:eval:selep}). Off hit and on hit are simulated block-cache hit rates without and with SeLeP prefetches, respectively. Miss cov. is eliminated misses over no-SeLeP misses; useless prefs are issued prefetches not later demanded.}
\label{tab:selep-random-paired-hit-rate}
\small
\begin{tabular}{@{}lrrrr@{}}
\toprule
Workload & Off hit & On hit & Miss cov. & Useless prefs \\
\midrule
LinkedList & 99.36\% & 99.99\% & 99.19\% & 1.60\% \\
Jacord & 93.11\% & 99.69\% & 95.45\% & 0.00\% \\
LittleX & 56.51\% & 58.35\% & 4.22\% & 71.20\% \\
JDrive & 65.45\% & 65.68\% & 0.68\% & 99.21\% \\
\bottomrule
\end{tabular}
\end{table}